\documentclass[12pt]{iopart}

\usepackage{iopams}  
\usepackage{bm}
\usepackage{here}
\usepackage{graphicx}

\begin{document}

\title[Supercurrent and EMF by Berry connection]{Supercurrent and Electromotive force generations by the Berry connection from many-body wave functions}

\author{Hiroyasu Koizumi}

\address{Division of Quantum Condensed Matter Physics, Center for Computational Sciences, University of Tsukuba, Tsukuba, Ibaraki 305-8577, Japan}
\ead{koizumi.hiroyasu.fn@u.tsukuba.ac.jp}
\vspace{10pt}
\begin{indented}
\item[]January  2023
\end{indented}

\begin{abstract}
The velocity field composed of the electromagnetic field vector potential and the Berry connection from many-body wave functions explains supercurrent generation, Faraday's law for the electromotive force (EMF) generation, and other EMF generations whose origins are not electromagnetism. An example calculation for the EMF from the Berry connection is performed using 
a model for the cuprate superconductivity.
\end{abstract}

%
%
%
%
%

\section{Introduction}

 The Berry phase first discovered in the context of the adiabatic approximation now prevails in various fields of physics \cite{Berry,Geometric}.
 In particular, it is now an indispensable mathematical tool to detect topological
 defects in quantum wave functions \cite{Ryu2016}. 
 Recently, the Berry connection from many-body wave functions was defined and its usefulness to calculate supercurrent 
 is demonstrated \cite{koizumi2022}. A salient feature of such a formalism is that it provides a vector potential directly related to the velocity field for electric current.
 In the present work, we consider the supercurrent and electromotive force (EMF) generations based on the same formalism \cite{koizumi2022,koizumi2022b}.
 
The EMF is expressed using a non-irrotational `electric field', ${\bf E}_{\rm irrot}$, whose origin may not be a real electric field. It is defined as
\begin{eqnarray}
{\cal E}=\oint_C{\bf E}_{\rm non-irrot} \cdot d{\bf r}
\end{eqnarray}
where $C$ is a closed electric circuit. This EMF appears due to various causes, such as chemical reactions in batters or
temperature differences in metals.

One of the important EMF generation mechanisms is the Faraday's law of magnetic induction. It is expressed as a total time-derivative of a magnetic flux of the magnetic field ${\bf B}$
\begin{eqnarray}
{\cal E}=-{d \over {dt}} \int_S {\bf B} \cdot d{\bf S}
\label{eq1a}
\end{eqnarray}
where $S$ is a surface whose circumference is $C$.
This EMF formula is often called the ``flux rule'', since $\int_S {\bf B} \cdot d{\bf S}$ is the magnetic flux through the surface $S$; it has been claimed curious since it is composed of two different fundamental equations in classical theory \cite{FeynmanII}, i.e., the Faraday's law of induction and the Lorentz force.
The curiosity is increased by the fact that one of them is
an equation for fields only, and the other includes particles and is an equation for a force on a particle. 

This peculiarity disappears in quantum theory 
using the vector potential ${\bf A}$ that is more fundamental than the magnetic field ${\bf B}$ \cite{AB1959,Tonomura1982,Tonomura1986}, and the wave function makes the velocity of a particle a velocity field \cite{London1950}.
Then, the two contributions in the ``flux rule'' are connected by the duality
that a $U(1)$ phase factor added on a wave function describes
a whole system motion, and also plays the role of the vector potential when it is transferred into the Hamiltonian \cite{FluxRule}.

In the present work, we extend the above vector potential and velocity field approach for the electric current generation
to cases where the vector potential of the Berry connection from many-body wave functions appears \cite{koizumi2022}.
We show that the EMF generation other than the electromagnetic field origin, such as those due to chemical reactions or temperature gradients can be expressed by it.

The organization of the present work is as follows:
we explain the velocity field appearing from the Berry connection from many-body wave functions in Section~\ref{Sec2}. We 
reexamine the Faraday's EMF generation formula using the velocity field from the electromagnetic vector potential in Section~\ref{Sec3}.
We examine the EMF generation by the Berry connection in Section~\ref{Sec4}, and an example calculation is performed for the Nernst effect in Section~\ref{Sec5}.
 Lastly, we conclude the present work by mentioning implications of the present new theory in Section~\ref{Sec6}.

\section{The velocity field from the Berry connection form many-body wave functions and supercurrent generation}
\label{Sec2}

The key ingredient in the present work is the
Berry connection from many-body wave functions for electrons given by
\begin{eqnarray}
{\bf A}^{\rm MB}_{\Psi}({\bf r})\!=\!
{ 1\over {\hbar \rho({\bf r})}} {\rm Re} \left\{
 \int d\sigma_1  d{\bf x}_{2}  \cdots d{\bf x}_{N}
 \Psi^{\ast}({\bf r}, \sigma_1, \cdots, {\bf x}_{N})
  (-i \hbar \nabla )
\Psi({\bf r}, \sigma_1, \cdots, {\bf x}_{N}) \right\}
\nonumber
\\
\label{Afic}
\end{eqnarray}
where $N$ is the total number of electrons in the system, 
`$\rm{Re}$' denotes the real part, $\Psi$ is the total wave function, ${\bf x}_i$ collectively stands for the coordinate ${\bf r}_i$ and the spin $\sigma_i$ of the $i$th electron, $-i \hbar \nabla$ is the Schr\"{o}dinger's momentum operator for the coordinate vector ${\bf r}$, and $\rho({\bf r})$ is the number
density calculated from $\Psi$. This Berry connection is obtained by regarding ${\bf r}$ as the ``adiabatic parameter''\cite{Berry}. 

Let us consider the electron system whose kinetic energy operator in the Schr\"{o}dinger representation is given by
\begin{eqnarray}
\hat{T}=- \sum_{j=1}^N
{\hbar ^2 \over {2 m_e}}\nabla_j^2
\label{eq4}
\end{eqnarray}
where $m_e$ is the electron mass.

For convenience, we also use the following $\chi$ defined as
  \begin{eqnarray}
{ {\chi({\bf r})}}= - 2\int^{{\bf r}}_0 {\bf A}_{\Psi}^{\rm MB}({\bf r}') \cdot d{\bf r}' 
\end{eqnarray}
and express the many-electron wave function $\Psi$ as
\begin{eqnarray}
\Psi({\bf x}_1, \cdots, {\bf x}_N)=\exp \left(  -{i \over 2} \sum_{j=1}^{N} \chi({\bf r}_j)\right)
\Psi_0({\bf x}_1, \cdots, {\bf x}_N)
\label{single-valued}
\end{eqnarray}
Then, $\Psi_0=\Psi \exp \left({i \over 2} \sum_{j=1}^{N} \chi({\bf r}_j)\right)$ is a currentless wave function for the current operator associated with $\hat{T}$ in Eq.~(\ref{eq4}) since the contribution from $\Psi$ and that from $\exp \left({i \over 2} \sum_{j=1}^{N} \chi({\bf r}_j)\right)$ cancel out.
In other words, a wave function is given as a product of a currentless one, $\Psi_0$, and the factor for the current $\exp \left(  -{i \over 2} \sum_{j=1}^{N} \chi({\bf r}_j)\right)$. The total wave function $\Psi$ must be a single-valued function of coordinates. This makes $\chi$ as an angular variable that satisfies some periodicity. This periodicity gives rise to non-trivial topological integer as will be explained, shortly.

When electromagnetic field is included, the kinetic energy operator becomes
\begin{eqnarray}
\hat{T}'= \sum_{j=1}^N
{1 \over {2 m_e}}(-i \hbar \nabla_j-q{\bf A})^2
\end{eqnarray}
where $q=-e$ is the electron charge, and ${\bf A}$ is the electromagnetic field vector potential. The magnetic field is given by ${\bf B}=\nabla \times {\bf A}$.

In the following, we will use the same expression, $\Psi$, for the total wave function.
Then, the current density for $\Psi$ is given by
\begin{eqnarray}
{\bf j}=-e \rho {\bf v}
\label{eq11}
\end{eqnarray}
with the velocity field ${\bf v}$ given by
\begin{eqnarray}
{\bf v}
&=&{e \over m_e}\left({\bf A}-{\hbar \over {2e}}\nabla \chi \right)
\nonumber
\\
&=&{e \over m_e}{\bf A}+{\hbar \over {m_e}}{\bf A}_{\Psi}^{\rm MB}
\label{eq12}
\end{eqnarray}

The current density in Eq.~(\ref{eq11}) is known to give rise to the Meissner effect if it is a stable one
due to the fact that it explicitly depends on ${\bf A}$ \cite{London1950}.
For the stable current case, $\nabla \chi$ compensates the gauge ambiguity in ${\bf A}$ and makes ${\bf v}$
in Eq.~(\ref{eq12}) gauge invariant.

If the Meissner effect is realize, the magnetic filed is expelled from the bulk of a superconductor \cite{London1950}.
Then, the flux quantization is observed for magnetic flux through a loop $C$ that goes through the bulk of a ring-shaped superconductor
\begin{eqnarray}
\int_S {\bf B}\cdot d{\bf S}&=&\oint_C {\bf A}\cdot d{\bf r}
\nonumber
\\
&=&{\hbar \over {2e}} \oint_C \nabla \chi \cdot d{\bf r}
\nonumber
\\
&=&{h \over {2e}} w_C[\chi]
\end{eqnarray}
where $w_C[\chi]$ is the topological integer `winding number' defined by
\begin{eqnarray}
 w_C[\chi]={ 1 \over {2 \pi}} \oint_C \nabla \chi \cdot d{\bf r}
\end{eqnarray}

According to Eq.~(\ref{eq12}), the presence of non-zero $w_C[\chi]$ means the existence of the stable velocity field that satisfies
\begin{eqnarray}
\oint_C {\bf v}\cdot d{\bf r}={h \over {2m_e}} w_C[\chi]
\label{eq14}
\end{eqnarray}

In superconductors, the quantized flux persists. This means
that the condition
\begin{eqnarray}
{d \over {dt}} w_C[\chi]=0
\label{eq15}
\end{eqnarray}
is realized.

In normal metals, the time-derivative of the velocity field is often expressed as
\begin{eqnarray}
{{d{\bf v}} \over {dt}}=-{1 \over \tau}{\bf v}
\end{eqnarray}
using a relaxation time approximation, where $\tau$ is the relaxation time.

Combination of this with Eq.~(\ref{eq14}) yields
\begin{eqnarray}
\tau {d \over {dt}}w_C[\chi] =-w_C[\chi]
\label{eq17}
\end{eqnarray}

If the condition in Eq.~(\ref{eq15}) with nonzero $w_C[\chi]$ is realized, Eq.~(\ref{eq17}) means that 
$\tau$ must be $\infty$, i.e., an infinite conductivity, or zero resistivity is realized.

\section{The vorticity field from the vector potential ${\bf A}$ and Faraday's flux rule}
\label{Sec3}

In this section, we consider the case where non-trivial ${\bf A}_{\Psi}^{\rm MB}$ is absent.
When ${\bf A}_{\Psi}^{\rm MB}$ is trivial, it satisfies 
\begin{eqnarray}
\nabla \times {\bf A}_{\Psi}^{\rm MB}=0
\end{eqnarray}
Thus, by applying $\nabla \times $ on the both sides of Eq.~(\ref{eq12})
\begin{eqnarray}
\nabla \times {\bf v}={e \over m_e} {\bf B}
\label{eq2-14}
\end{eqnarray}
is obtained.

Taking the total time-derivative of the above yields
\begin{eqnarray}
\nabla \times {{d{\bf v}} \over {dt}}={e \over m_e} \partial_t{\bf B}+{e \over m_e}({\bf v} \cdot \nabla){\bf B}
\label{eq2-15}
\end{eqnarray}
where the total time-derivative of the field ${\bf B}$ is the Eulerian time-derivative given by
\begin{eqnarray}
 {{d{\bf B}} \over {dt}}= \partial_t{\bf B}+({\bf v} \cdot \nabla){\bf B}
\end{eqnarray}

Integrating Eq.~(\ref{eq2-15}) over the surface $S$, we have
\begin{eqnarray}
\oint_C {{d{\bf v}} \over {dt}}\cdot d{\bf r}={e \over m_e}\int_S \partial_t{\bf B}\cdot d{\bf S}+{e \over m_e}
\int_S({\bf v} \cdot \nabla){\bf B}\cdot d{\bf S}
\end{eqnarray}
where the Stokes theorem is used to convert the surface integral to the line integral.

Noting that the electromotive force for an electron is given by
\begin{eqnarray}
{\cal E}={1 \over {-e}}\oint_C {{d (m_e{\bf v})} \over {dt}}\cdot d{\bf r}
\end{eqnarray}
where $-e$ is the electron charge and $m_e$ is the electron mass,
the following relation is obtained
\begin{eqnarray}
{\cal E}=-\int_S \partial_t{\bf B}\cdot d{\bf S}-
\int_S({\bf v} \cdot \nabla){\bf B}\cdot d{\bf S}
\end{eqnarray}
This is equal to the Faraday's formula in Eq.~(\ref{eq1a}).

In the situation where the circuit $C$ moves with a constant velocity ${\bf v}_0$, we have the following relation
\begin{eqnarray}
({\bf v}_0 \cdot \nabla){\bf B}&=&\nabla \times ({\bf B}\times {\bf v}_0)+{\bf v}_0(\nabla \cdot {\bf B})
\nonumber
\\
&=&\nabla \times ({\bf B}\times {\bf v}_0)
\end{eqnarray}
due to the fact that ${\bf B}$ satisfies $\nabla \cdot {\bf B}=0$ \cite{Jackson}.

As a consequence, the well-known EMF formula
\begin{eqnarray}
{\cal E}=-\int_S \partial_t{\bf B}\cdot d{\bf S}+
\oint_C({\bf v}_0 \times {\bf B})\cdot d{\bf r}
\label{eq24}
\end{eqnarray}
is obtained. The first term in it is attributed to the Faraday's law of induction, and the second to the Lorentz force. This formula is composed of two different fundamental equations in classical theory \cite{FeynmanII}.
However, in the quantum mechanical formalism, two contributions stem from
a single relation in Eq.~(\ref{eq12}). 

\section{The EMF generation by the Berry connection}
\label{Sec4}

The velocity field in Eq.~(\ref{eq12}) contains the vector potential ${\bf A}_{\Psi}^{\rm MB}$ in addition to the electromagnetic vector potential ${\bf A}$. Just like ${\bf A}$, ${\bf A}_{\Psi}^{\rm MB}$ will also give rise to the EMF. 

We now consider a general case where the Berry connection arises from a set of states $\{ \Psi_j \}$ and given by
\begin{eqnarray}
{\bf A}^{\rm MB}=\sum_j p_j {\bf A}_{\Psi_j}^{\rm MB}
\end{eqnarray}
where $p_j$'s are probabilities satisfy
\begin{eqnarray}
\sum_j p_j=1
\end{eqnarray}
and ${\bf A}_{\Psi_j}^{\rm MB}$ is obtained from Eq.~(\ref{Afic}) by replacing $\Psi$ with $\Psi_j$.

We express ${\bf A}^{\rm MB}$ using the following density matrix
\begin{eqnarray}
\hat{d}=\sum_j p_j|\Psi_j \rangle \langle \Psi_j |
\end{eqnarray}
where the operator $\hat{\bf A}^{\rm MB}$ is defined through the relation
\begin{eqnarray}
\langle \Psi_j |\hat{\bf A}^{\rm MB} |\Psi_j \rangle ={\bf A}_{\Psi_j}^{\rm MB}
\end{eqnarray}
From now on, we allow the time-dependence in $\Psi_j$. When $\Psi_j$ is time-dependent, ${\bf A}_{\Psi_j}^{\rm MB}$ 
is also time-dependent.
The distribution probability $p_j$ can be also 
time and coordinate dependent.

Using the density operator $\hat{d}$ and the operator $\hat{\bf A}^{\rm MB}$, the vector potential from the Berry connection is given by
\begin{eqnarray}
{\bf A}^{\rm MB}={\rm tr} \left(\hat{d}\hat{\bf A}^{\rm MB}\right)
\end{eqnarray}

We define ${\bf B}^{\rm MB}$ by
\begin{eqnarray}
{\bf B}^{\rm MB}=\nabla \times {\bf A}^{\rm MB}
\end{eqnarray}

Then, the EMF from the Berry connection is given by
\begin{eqnarray}
{\cal E}^{\rm MB}=-{ \hbar \over e}\int_S \partial_t{\bf B}^{\rm MB}\cdot d{\bf S}-{ \hbar \over e}
\int_S({\bf v}\cdot \nabla) {\bf B}^{\rm MB}\cdot d{\bf S}
\label{eq28}
\end{eqnarray}
The first term in the right hand side can arise from the time-dependence of $p_j$.
This means that if $p_j$ varies with time due to chemical reactions, photo excitations, or etc. it will give rise to the EMF.
The second term will arise if the temperature depends on the coordinate, $T({\bf r})$, and $p_j$ contains the Boltzmann factor $\exp ( -{E_j \over {k_B T({\bf r})}})$, where $E_j$ is the energy for the state $\Psi_j$. It also arises when $p_j$ depends on the coordinate due, for example,  to the concentration gradient of chemical spices. 

Now we consider the case where the circuit moves with a constant vector ${\bf v}_0$.
The circuit in this case should be regarded as a region of the system which flows due to the flow existing in the system.
Such a motion may arise from a temperature gradient or concentration gradient in the system.
In this case, we have the following relation,
\begin{eqnarray}
({\bf v}\cdot \nabla) {\bf B}^{\rm MB}=-\nabla \times ({\bf v}_0 \times {\bf B}^{\rm MB})
\end{eqnarray}
due to the fact that $\nabla \cdot {\bf B}^{\rm MB}=\nabla \cdot (\nabla \times {\bf A}^{\rm MB})=0$.

The equation (\ref{eq28}) can be cast into the following form
\begin{eqnarray}
{\cal E}^{\rm MB}=-{ \hbar \over e}\oint_C \left[
\partial_t{\bf A}^{\rm MB}-{\bf v}_0 \times (\nabla \times {\bf A}^{\rm MB})
\right] \cdot d{\bf r}
\label{eq30}
\end{eqnarray}
that only contains ${\bf A}^{\rm MB}$.
However, the above formula may not be convenient to use due to the fact that
${\bf A}^{\rm MB}$ contains topological singularities.
A convenient one may be the following 
\begin{eqnarray}
{\cal E}^{\rm MB}=-{ \hbar \over e} {d \over {dt}} \int_S {\bf B}^{\rm MB} \cdot d{\bf S}
\label{eq31}
\end{eqnarray}
where ${\bf B}$ in the Faraday's law in Eq.~(\ref{eq1a}) 
is replaced by ${\bf B}^{\rm MB}$. 

\section{Nernst effect}
\label{Sec5}

In this section, we examine the Nernst effect observed in cuprate superconductors \cite{Nernst,Nernst2005,Nernst2}.
We examine this phenomenon using Eq.~(\ref{eq31}).
A theory of superconductivity in the cuprate predicts the appearance of spin-vortices in the CuO$_2$ plane around doped holes that become small polarons \cite{koizumi2,Koizumi2011,Hidekata2011}. The spin-vortices generate the vector potential
\begin{eqnarray}
{\bf A}^{\rm MB} =-{1 \over 2} \nabla \chi
\end{eqnarray}
where $\chi$ is an angular variable with period $2\pi$. This angular variable appears due to the requirement that
the wave function to be a single-valued function of coordinates in the situation where itinerant motion of electrons 
around the small polaron hole is a spin-twisting one.

We can decompose $\chi$ as a sum over spin-vortices
\begin{eqnarray}
\chi =\sum_{j=1}^{N_h} \chi_j
\end{eqnarray}
where $\chi_j$ is a contribution form the $j$th small polaron hole, and $N_h$ is the total number of holes that become small polarons.

Each $\chi_j$ is characterized by its winding number
\begin{eqnarray}
w_j={1 \over {2 \pi}}\oint_{C_j}\nabla \chi_j \cdot d{\bf r}
\end{eqnarray}
where $C_j$ is a loop that only encircles the center of the $j$th spin-vortex. We can assume $w_j$ to be $+1$ or $-1$;
only odd integers are allowed due to the spin-twisting motion. The numbers $\pm 1$ are favorable from the energetic point of view.

\begin{figure}[H]
\begin{center}
\includegraphics[width=10.0cm]{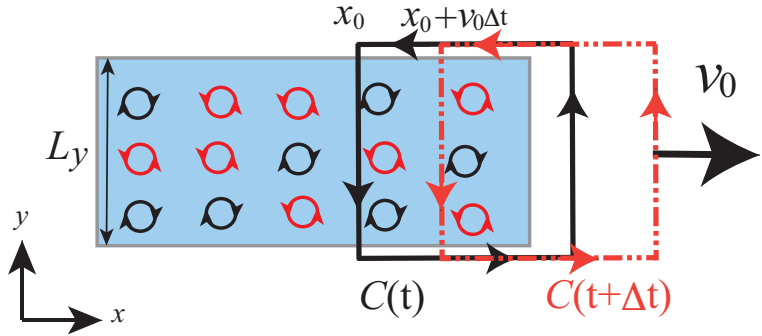}
\end{center}
\caption{A schematic picture for the EMF appearing from the Berry connection generated by spin-vortices.
The Berry connection creates the vector potential proportional to $\nabla \chi$, which creates vortices (loop currents) denoted by
circles with arrows. We consider two loops $C(t)$ and $C(t+\Delta t)$, where $t$ and $t+\Delta t$ denote two times with interval $\Delta t$. The loop moves with velocity $v_0$ in the $x$-direction due to the temperature gradient in that direction. A constant magnetic field is applied in the $z$-direction. A voltage is generated across the $y$-direction.
The sample exists $0 \leq y \leq L_y$. The left edge of the loop at time $t$ is $x_0$ and that at time $t+\Delta t$ is $x_0+v_0 \Delta t$.}
\label{FigNernst}
\end{figure}

Let us consider the situation depicted in Fig.~\ref{FigNernst}. We neglect the contribution from ${\bf A}$ assuming that it is small. The EMF generated across the sample in the $y$-direction is given by
\begin{eqnarray}
{\cal E}^{\rm MB}&=&
-{ \hbar \over e} {1 \over {\Delta t}} \left[
\int_{S(t+\Delta t)} {\bf B}^{\rm MB} \cdot d{\bf S}
-
\int_{S(t)} {\bf B}^{\rm MB} \cdot d{\bf S}
\right]
\nonumber
\\
&=&
-{ \hbar \over e} {1 \over {\Delta t}} \left[
\oint_{C(t+\Delta t)} {\bf A}^{\rm MB} \cdot d{\bf r}
-
\oint_{C(t)} {\bf A}^{\rm MB} \cdot d{\bf r}
\right]
\nonumber
\\
&=&
{ \hbar \over e} {1 \over {\Delta t}}
\oint_{\Delta C} {\bf A}^{\rm MB} \cdot d{\bf r}
\label{eq38}
\end{eqnarray}
where $S(t+\Delta t)$ and $S(t)$ are surfaces in the $xy$-plane with circumferences $C(t+\Delta t)$ and $C(t)$,
respectively; $\Delta C$ is the loop encircling the area $x_0 \leq x \leq x_0+v_0 \Delta t$, $0 \leq y \leq L_y$, with
the counterclockwise direction.

We approximate $\oint_{\Delta C} {\bf A}^{\rm MB} \cdot d{\bf r}$ by
\begin{eqnarray}
\oint_{\Delta C} {\bf A}^{\rm MB} \cdot d{\bf r}&=&-{1 \over 2} \oint_{\Delta C} \nabla \chi \cdot d{\bf r}
\nonumber
\\
&\approx&
-{1 \over 2} 2 \pi (n_m-n_a)L_y v_0 \Delta t
\label{eq39}
\end{eqnarray}
where $n_m$ and $n_a$ are average densities of $w_j=1$ (`meron') and $w_j=-1$ (`antimeron') vortices, respectively.
Thus, $n_mL_y v_0 \Delta t$ and $n_aL_y v_0 \Delta t$ are expected numbers of $w_j=1$ and $w_j=-1$ vortices within the loop ${\Delta C}$, respectively.

From Eqs.~(\ref{eq38}) and (\ref{eq38}), the approximate ${\cal E}^{\rm MB}$ is given by
\begin{eqnarray}
{\cal E}^{\rm MB} \approx {{h v_0} \over {2e}} (n_a-n_m)L_y 
\end{eqnarray}

Thus, the electric field generated by ${\cal E}^{\rm MB}$ in the $y$-direction is given by
\begin{eqnarray}
E_y \approx {{h v_0} \over {2e}} (n_a-n_m)
\end{eqnarray}

In our previous work, $n_a$ is denoted as $n_d$ indicting that it yields a diamagnetic current, and $n_m$ as $n_p$ indicting that it yields a paramagnetic current \cite{Koizumi2011,Hidekata2011}. 
Using $n_d$ and $n_p$, the Nernst signal is obtained as
\begin{eqnarray}
e_N={{E_y} \over {|\partial_x T|}}={{h v_0 (n_d-n_p)} \over {2e|\partial_x T|}} 
\label{eqNernst2}
\end{eqnarray}
The same formula was obtained previously for the situation where spin-vortices move by the temperature gradient \cite{Koizumi2011,Hidekata2011}.
Here, the situation is different; the spin-vortices do not move, but the electron system affected by $\nabla \chi$ moves. Considering that the small polaron movement is negligible at low temperature, the present situation is more realistic than the previous one. The temperature dependence is the same as the one that qualitatively explains the experimental result \cite{Hidekata2011}.

Note that experiments indicating the presence of loop currents different from
ordinary Abrikosov vortices \cite{Abrikosov} in the cuprate \cite{Kerr1,Kerr2}. 
The present result indicates that the observed Nernst
can be explained by the presence of spin-vortex-induced loop currents.

\section{Concluding remarks}
\label{Sec6}

Since the EMF by the Berry connection is not the electromagnetic field origin, it may be more appropriate to call it the Berry-connection motive force (BCMF) given by
\begin{eqnarray}
{\cal F}^{\rm  BMF}=-e {\cal E}^{\rm  MB}=
\hbar {d \over {dt}} \int_S {\bf B}^{\rm MB} \cdot d{\bf S}
\end{eqnarray}

The BCMF will arise from quantum mechanical dynamics of particles other than electrons; for example, from proton dynamics, through chemical reactions. The non-trivial Berry phase effect has been predicted \cite{Mead79}, and  observed in the hydrogen transfer reactions \cite{Yuan1289}. Quantum mechanical effects are important in such reactions due to the relatively light mass of protons \cite{Schatz1975,Wyatt1975}.
It is known that the EMF generated by the proton pumps is a very important chemical process in biological systems,
and the Berry-connection motive force may play some roles in the working of the proton pumps. It may be also useful to invent high performance batteries.

\section*{References}

\providecommand{\newblock}{}


\end{document}